\documentclass[english]{article}
\usepackage[latin9]{inputenc}
\usepackage{color}
\usepackage{amssymb}

\makeatletter
\newcommand{\lyxaddress}[1]{
\par {\raggedright #1
\vspace{1.4em}
\noindent\par}
}

\makeatother

\usepackage{babel}

\begin{document}

\title{\textbf{Effective temperature for black holes}}

\author{\textbf{Christian Corda}}

\maketitle

\lyxaddress{\begin{center}
Institute for Theoretical Physics and Mathematics Einstein-Galilei,
Via Santa Gonda 14, 59100 Prato, Italy
\par\end{center}}

\lyxaddress{\begin{center}
\textit{E-mail addresses:} \textcolor{blue}{cordac.galilei@gmail.com} 
\par\end{center}}
\begin{abstract}
The physical interpretation of black hole's quasinormal modes is fundamental
for realizing unitary quantum gravity theory as black holes are considered
theoretical laboratories for testing models of such an ultimate theory
and their quasinormal modes are natural candidates for an interpretation
in terms of quantum levels. 

The spectrum of black hole's quasinormal modes can be re-analysed
by introducing a black hole's \emph{effective temperature }which takes
into account the fact that, as shown by Parikh and Wilczek, the radiation
spectrum cannot be strictly thermal. This issue changes in a fundamental
way the physical understanding of such a spectrum and enables a re-examination
of various results in the literature which realizes important modifies
on quantum physics of black holes. In particular, the formula of the
horizon's area quantization and the number of quanta of area result
modified becoming functions of the quantum {}``overtone'' number
$n$. Consequently, the famous formula of Bekenstein-Hawking entropy,
its sub-leading corrections and the number of microstates are also
modified. Black hole's entropy results a function of the quantum overtone
number too. 

We emphasize that this is the first time that black hole's entropy
is directly connected with a quantum number.

Previous results in the literature are re-obtained in the limit $n\rightarrow\infty$. \end{abstract}
\begin{description}
\item [{Keywords:}] Black hole, entropy, Hawking radiation, effective temperature.
\item [{PACS}] \textbf{numbers}: 04.70.-s, 04.70.Dy.
\end{description}
Hawking radiation \cite{key-1} can be visualized as particles that
have tunnelled across the black hole's horizon \cite{key-2,key-3}.
In \cite{key-2,key-3} Parikh and Wilczek showed that the barrier
depends on the tunnelling particle itself. Parikh released an intriguing
physical interpretation of this fundamental issue by discussing the
existence of a tunnel through the black hole's horizon \cite{key-2}.
By implementing energy conservation, the black hole contracts during
the process of radiation. Thus, the horizon recedes from its original
radius to a new, smaller radius \cite{key-2}. Hence, the radiation
spectrum cannot be strictly thermal. The correction to the thermal
spectrum has profound implications for realizing unitary quantum gravity
theory and for the black hole's information puzzle \cite{key-2,key-3}.
In fact, black holes are considered theoretical laboratories for developing
unitary quantum gravity theory and their quasinormal modes are the
best candidates for an interpretation in terms of quantum levels. 

In this work we re-analyse the spectrum of black hole's quasinormal
modes by taking into account the issue that the radiation spectrum
is not strictly thermal. 

Working with $G=c=k_{B}=\hbar=\frac{1}{4\pi\epsilon_{0}}=1$ (Planck
units) the probability of tunnelling takes the form \cite{key-1,key-2,key-3} 

\begin{equation}
\Gamma\sim\exp(-\frac{\omega}{T_{H}}),\label{eq: hawking probability}\end{equation}

where $T_{H}\equiv\frac{1}{8\pi M}$ is the Hawking temperature and
$\omega$ the energy-frequency of the emitted radiation.

The remarkable correction by Parikh and Wilczek, due by an exact calculation
of the action for a tunnelling spherically symmetric particle, yields
\cite{key-2,key-3}

\begin{equation}
\Gamma\sim\exp[-\frac{\omega}{T_{H}}(1-\frac{\omega}{2M})].\label{eq: Parikh Correction}\end{equation}

This result has also taken into account the conservation of energy
and this enables a correction, the additional term $\frac{\omega}{2M}$
\cite{key-2}. If we introduce the \emph{effective temperature} (which
depends from the energy-frequency of the emitted radiation) 

\begin{equation}
T_{E}(\omega)\equiv\frac{2M}{2M-\omega}T_{H}=\frac{1}{4\pi(2M-\omega)},\label{eq: Corda Temperature}\end{equation}

Eq. (\ref{eq: Parikh Correction}) can be rewritten in Boltzmann-like
form 

\begin{equation}
\Gamma\sim\exp[-\beta_{E}(\omega)\omega]=\exp(-\frac{\omega}{T_{E}(\omega)}),\label{eq: Corda Probability}\end{equation}

where $\beta_{E}(\omega)\equiv\frac{1}{T_{E}(\omega)}$ and $\exp[-\beta_{E}(\omega)\omega]$
is the \emph{effective Boltzmann factor} appropriate for an object
with inverse effective temperature $T_{E}(\omega).$ The ratio $\frac{T_{E}(\omega)}{T_{H}}=\frac{2M}{2M-\omega}$
represents the deviation of the radiation spectrum of a black hole
from the strictly thermal feature. In other terms, as the correction
in \cite{key-2,key-3} implies that a black hole does not strictly
emit like a black body, the effective temperature represents the temperature
of a black body that would emit the same total amount of radiation. 

Now, we apply the introduction of the effective temperature $T_{E}(\omega)$
to the analysis of the spectrum of black hole's quasinormal modes. 

For Schwarzschild black holes, the quasinormal mode frequencies are
usually labelled as $\omega_{nl},$ where $l$ is the angular momentum
quantum number \cite{key-4,key-5}. For each $l$ ($l$$\geq2$ for
gravitational perturbations), we have a countable infinity of quasinormal
modes, labelled by the {}``overtone'' number $n$ ($n=1,2,...$)
\cite{key-5}. For large $n$ the frequencies of quasinormal modes
for the Schwarzschild black hole become independent of $l$ having
the structure \cite{key-4,key-5} \begin{equation}
\begin{array}{c}
\omega_{n}=\ln3\times T_{H}+2\pi i(n+\frac{1}{2})\times T_{H}+\mathcal{O}(n^{-\frac{1}{2}})=\\
\\=\frac{\ln3}{8\pi M}+\frac{2\pi i}{8\pi M}(n+\frac{1}{2})+\mathcal{O}(n^{-\frac{1}{2}}).\end{array}\label{eq: quasinormal modes}\end{equation}

This result was originally obtained numerically in \cite{key-6,key-7},
while an analytic proof was given in \cite{key-8,key-9}. 

In any case, Eq. (\ref{eq: quasinormal modes}) is an approximation
as it has been derived with the assumption that the black hole radiation
spectrum is strictly thermal. To take into due account the deviation
from the thermal spectrum in \cite{key-2,key-3} one has to substitute
the Hawking temperature $T_{H}$ with the effective temperature $T_{E}$
in Eq. (\ref{eq: quasinormal modes}). In this way, the correct expression
for the frequencies of quasinormal modes for the Schwarzschild black
hole, which takes into account the important issue that the radiation
spectrum is not strictly thermal, is

\begin{equation}
\begin{array}{c}
\omega_{n}=\ln3\times T_{E}(\omega_{n})+2\pi i(n+\frac{1}{2})\times T_{E}(\omega_{n})+\mathcal{O}(n^{-\frac{1}{2}})=\\
\\=\frac{\ln3}{4\pi(2M-\omega_{n})}+\frac{2\pi i}{4\pi(2M-\omega_{n})}(n+\frac{1}{2})+\mathcal{O}(n^{-\frac{1}{2}}).\end{array}\label{eq: quasinormal modes corrected}\end{equation}

Let us explain this key point. The imaginary part of (\ref{eq: quasinormal modes})
is simple to understand \cite{key-9}. The quasinormal modes determine
the position of poles of a Green's function on the given background,
and the Euclidean black hole solution converges to a thermal circle
at infinity with the inverse temperature $\beta_{H}=\frac{1}{T_{H}}$
\cite{key-9}. Hence, it is not surprising that the spacing of the
poles in (\ref{eq: quasinormal modes}) coincides with the spacing
$2\pi iT_{H}$ expected for a thermal Green's function \cite{key-9}.
But, if we want to consider the deviation from the thermal spectrum
which has been found in \cite{key-2,key-3} it is natural to assume
that the Euclidean black hole solution converges to a \emph{non-thermal}
circle at infinity. Therefore, it is straightforward the substitution

\begin{equation}
\beta_{H}=\frac{1}{T_{H}}\rightarrow\beta_{E}(\omega)=\frac{1}{T_{E}(\omega)},\label{eq: sostituiamo}\end{equation}

which takes into account the deviation of the radiation spectrum of
a black hole from the strictly thermal feature. In this way, the spacing
of the poles in (\ref{eq: quasinormal modes corrected}) coincides
with the spacing 

\begin{equation}
2\pi iT_{E}(\omega)=2\pi iT_{H}(\frac{2M}{2M-\omega}),\label{eq: spacing}\end{equation}

expected for a non-thermal Green's function (a dependence from the
frequency is present).

On the other hand, one could be not satisfied of a similar classical
intuitive explanation to substitute Eq. (\ref{eq: quasinormal modes corrected})
for Eq. (\ref{eq: quasinormal modes}). Hence, we further release
a rigorous argument. We recall that quasinormal modes are frequencies
of the radial spin-j perturbations $\phi$ of the four-dimensional
Schwarzschild background which are governed by the following differential
equation \cite{key-8,key-9}

\begin{equation}
\left(-\frac{\partial^{2}}{\partial x^{2}}+V(x)-\omega^{2}\right)\phi.\label{eq: diff.}\end{equation}

This equation is treated as a Schrodinger equation with the Regge-Wheeler
potential ($j=2$ for gravitational perturbations) \cite{key-8,key-9}

\begin{equation}
V(x)=V\left[x(r)\right]=\left(1-\frac{2M}{r}\right)\left(\frac{l(l+1)}{r^{2}}-\frac{6M}{r^{3}}\right).\label{eq: Regge-Wheeler}\end{equation}

The Regge-Wheeler {}``tortoise'' coordinate $x$ is related to the
radial coordinate $r$ by \cite{key-8,key-9}

\begin{equation}
\begin{array}{c}
x=r+2M\ln\left(\frac{r}{2M}-1\right)\\
\\\frac{\partial}{\partial x}=\left(1-\frac{2M}{r}\right)\frac{\partial}{\partial r}.\end{array}\label{eq: tortoise}\end{equation}

By realizing a rigorous analytical calculation, in \cite{key-8} Motl
derived Eq. (\ref{eq: quasinormal modes}) starting from Eqs. (\ref{eq: diff.})
and (\ref{eq: Regge-Wheeler}) and satisfying purely outgoing boundary
conditions both at the horizon ($r=2M$) and in the asymptotic region
($r=\infty$). But, if we want to take into due account the conservation
of energy, we have to substitute the original black hole's mass $M$
in Eqs. (\ref{eq: diff.}) and (\ref{eq: Regge-Wheeler}) with an
\emph{effective mass} of the contracting black hole. In other words,
if $M$ is the initial mass of the black hole \emph{before} the emission,
and $M-\omega$ is the final mass of the hole \emph{after} the emission
\cite{key-3}, Eqs. (\ref{eq: Parikh Correction}) and (\ref{eq: Corda Temperature})
enable the introduction of the effective mass \[
M_{E}\equiv M-\frac{\omega}{2}\]

of the black hole \emph{during} the emission of the particle, i.e.
\emph{during} the contraction's phase of the black hole. Notice that
the introduced effective mass is a perfect average of the initial
and final masses. Then, Eqs. (\ref{eq: Regge-Wheeler}) and (\ref{eq: tortoise})
have to be substituted with the \emph{effective equations}

\begin{equation}
V(x)=V\left[x(r)\right]=\left(1-\frac{2M_{E}}{r}\right)\left(\frac{l(l+1)}{r^{2}}-\frac{6M_{E}}{r^{3}}\right)\label{eq: effettiva 1}\end{equation}

and \begin{equation}
\begin{array}{c}
x=r+2M_{E}\ln\left(\frac{r}{2M_{E}}-1\right)\\
\\\frac{\partial}{\partial x}=\left(1-\frac{2M_{E}}{r}\right)\frac{\partial}{\partial r}.\end{array}\label{eq: effettiva 2}\end{equation}

If one realizes step by step the same rigorous analytical calculation
in \cite{key-8}, but starting from Eqs. (\ref{eq: diff.}) and (\ref{eq: effettiva 1})
and satisfying purely outgoing boundary conditions both at the \emph{effective
horizon} ($r=2M_{E}$) and in the asymptotic region ($r=\infty$),
the final result will be, obviously and rigorously, Eq. (\ref{eq: quasinormal modes corrected}).

Now, we can proceed with our analysis.

Notice that in Eq. (\ref{eq: quasinormal modes corrected}) the frequency
$\omega_{n}$ is present in both of the left hand side and the right
hand side. One could solve this equation and write down an analytic
form for $\omega_{n}$ but we will see that this is not essential
for our goals.

In \cite{key-5} the spectrum of black hole's quasinormal modes has
been analysed in terms of superposition of damped oscillations, of
the form \begin{equation}
\exp(-i\omega_{I}t)[a\sin\omega_{R}t+b\cos\omega_{R}t]\label{eq: damped oscillations}\end{equation}

with a spectrum of complex frequencies $\omega=\omega_{R}+i\omega_{I}.$
A damped harmonic oscillator $\mu(t)$ is governed by the equation
\cite{key-5} \begin{equation}
\ddot{\mu}+K\dot{\mu}+\omega_{0}^{2}\mu=F(t),\label{eq: oscillatore}\end{equation}

where $K$ is the damping constant, $\omega_{0}$ the proper frequency
of the harmonic oscillator, and $F(t)$ an external force per unit
mass. If $F(t)\sim\delta(t),$ i.e. considering the response to a
Dirac delta function, the result for $\mu(t)$ is a superposition
of a term oscillating as $\exp(i\omega t)$ and of a term oscillating
as $\exp(-i\omega t)$, see \cite{key-5} for details. Then, the behavior
(\ref{eq: damped oscillations}) is reproduced by a damped harmonic
oscillator, through the identifications \cite{key-5}

\begin{equation}
\begin{array}{ccc}
\frac{K}{2}=\omega_{I}, &  & \sqrt{\omega_{0}^{2}-\frac{K}{4}^{2}}=\omega_{R},\end{array}\label{eq: identificazioni}\end{equation}

which gives \begin{equation}
\omega_{0}=\sqrt{\omega_{R}^{2}+\omega_{I}^{2}}.\label{eq: omega 0}\end{equation}

An important point emphasized in \cite{key-5} is that identification
$\omega_{0}=\omega_{R}$ is correct only in the approximation $\frac{K}{2}\ll\omega_{0},$
i.e. only for very long-lived modes. For a lot of black hole's quasinormal
modes, for example for highly excited modes, the opposite limit can
be correct. In \cite{key-5} this observation has been used to re-examine
some aspects of quantum physics of black holes that were discussed
in previous literature assuming that the relevant frequencies were
$(\omega_{R})_{n}$ than $(\omega_{0})_{n}$. Here, we further improve
the analysis by taking into account the important issue that the radiation
spectrum is not strictly thermal. Let us modify the analysis in \cite{key-5}.
By using the new expression (\ref{eq: quasinormal modes corrected})
for the frequencies of quasinormal modes, we define \begin{equation}
\begin{array}{ccc}
m_{0}\equiv\frac{\ln3}{4\pi[2M-(\omega_{0})_{n}]}, &  & p_{n}\equiv\frac{2\pi}{4\pi[2M-(\omega_{0})_{n}]}(n+\frac{1}{2}).\end{array}\label{eq: definizioni}\end{equation}

Then, Eq. (\ref{eq: omega 0}) can be rewritten in the enlightening
form \begin{equation}
(\omega_{0})_{n}=\sqrt{m_{0}^{2}+p_{n}^{2}}.\label{eq: enlightening}\end{equation}

These results improve Eqs. (8) and (9) in \cite{key-5} as the new
expression (\ref{eq: quasinormal modes corrected}) for the frequencies
of quasinormal modes takes into account that the radiation spectrum
is not strictly thermal. For highly excited modes 

\begin{equation}
(\omega_{0})_{n}\approx p_{n}=\frac{2\pi}{4\pi[2M-(\omega_{0})_{n}]}(n+\frac{1}{2}).\label{eq: non equalmente spaziati}\end{equation}
 Thus, differently from \cite{key-5}, levels are \emph{not} equally
spaced even for highly excited modes. Indeed, there are deviations
due to the non-strictly thermal behavior of the spectrum (black hole's
effective temperature depends on the energy level). 

Using Eq. (\ref{eq: definizioni}),  Eq. (\ref{eq: enlightening})
can be rewritten like

\begin{equation}
(\omega_{0})_{n}=\frac{1}{4\pi[2M-(\omega_{0})_{n}]}\sqrt{(\ln3)^{2}+4\pi^{2}(n+\frac{1}{2})^{2}},\label{eq: enlightening 2}\end{equation}

which is easily solved giving \begin{equation}
(\omega_{0})_{n}=M\pm\sqrt{M^{2}-\frac{1}{4\pi}\sqrt{(\ln3)^{2}+4\pi^{2}(n+\frac{1}{2})^{2}}}.\label{eq: radici}\end{equation}

Clearly, only the solution $(\omega_{0})_{n}\ll M$ has physical meaning,
i.e. the one with the sign minus in the right hand side

\begin{equation}
(\omega_{0})_{n}=M-\sqrt{M^{2}-\frac{1}{4\pi}\sqrt{(\ln3)^{2}+4\pi^{2}(n+\frac{1}{2})^{2}}}.\label{eq: radice fisica}\end{equation}

The interpretation is of a particle quantized with anti-periodic boundary
conditions on a circle of length \begin{equation}
L=\frac{1}{T_{E}(\omega_{n})}=4\pi\left(M+\sqrt{M^{2}-\frac{1}{4\pi}\sqrt{(\ln3)^{2}+4\pi^{2}(n+\frac{1}{2})^{2}}}\right),\label{eq: lunghezza cerchio}\end{equation}
 i.e. the length of the circle depends from the overtone number $n.$
In \cite{key-5} Maggiore found a particle quantized with anti-periodic
boundary conditions on a circle of length $L=8\pi M.$ Our correction
takes into account the conservation of energy, i.e. the additional
term $\frac{\omega}{2M}$ in Eq. (\ref{eq: Parikh Correction}). 

Now, let us see various important consequences of the above approach
on the quantum physics of black holes starting by the \emph{area quantization}. 

Bekenstein \cite{key-10} showed that the area quantum of the Schwarzschild
black hole is $\triangle A=8\pi$ (we recall that the \emph{Planck
distance} $l_{p}=1.616\times10^{-33}\mbox{ }cm$ is equal to one in
Planck units). By using properties of the spectrum of Schwarzschild
black hole quasinormal modes a different numerical coefficient has
been found by Hod in \cite{key-11}. The analysis in \cite{key-11}
started by the observation that, as for the Schwarzschild black hole
the \emph{horizon area} $A$ is related to the mass through the relation
$A=16\pi M^{2},$ a variation $\triangle M$ in the mass generates
a variation

\begin{equation}
\triangle A=32\pi M\triangle M\label{eq: variazione area}\end{equation}

in the area. By considering a transition from an unexcited black hole
to a black hole with very large $n$, Hod assumed \emph{Bohr's correspondence
principle} to be valid for large $n$ and enabled a semiclassical
description even in absence of a full unitary quantum gravity theory.
Thus, from Eq. (\ref{eq: quasinormal modes}), the minimum quantum
which can be absorbed in the transition is $\triangle M=\omega=\frac{\ln3}{8\pi M}.$
This gives $\triangle A=4\ln3.$ The presence of the numerical factor
$4\ln3$ stimulated possible connections with loop quantum gravity
\cite{key-12}. By using Eq. (\ref{eq: quasinormal modes corrected})
than Eq. (\ref{eq: quasinormal modes}), Hod's result can be improved.
We get

\begin{equation}
\triangle M=\omega=\frac{\ln3}{4\pi(2M-\omega)},\label{eq: Hod improved}\end{equation}

which is easily solved giving

\begin{equation}
\triangle M=M\pm\sqrt{M^{2}-\frac{1}{4\pi}\ln3}.\label{eq: Hod improved solution}\end{equation}

Even in this case, only the solution $\triangle M\ll M$ has physical
meaning, i.e. the one with the sign minus in the right hand side,
\begin{equation}
\triangle M=M-\sqrt{M^{2}-\frac{1}{4\pi}\ln3}.\label{eq: Hod improved fisica}\end{equation}

Again, the modify takes into account the conservation of energy, i.e.
the additional term $\frac{\omega}{2M}$ in Eq. (\ref{eq: Parikh Correction})
that represents the deviation of the radiation spectrum of a black
hole from the strictly thermal feature. By using Eq. (\ref{eq: variazione area})
we get

\begin{equation}
\triangle A=32\pi M(M-\sqrt{M^{2}-\frac{1}{4\pi}\ln3}).\label{eq: Hod corretto}\end{equation}

Criticism on Hod's conjecture were discussed in \cite{key-5}. The
main point is that Bohr's correspondence principle strictly holds
only for transitions from $n$ to $n'$ where both $n,n'\gg1.$ Thus,
Maggiore \cite{key-5} suggested that $(\omega_{0})_{n}$ should be
used than $(\omega_{R})_{n}$, by obtaining the original Bekenstein's
result, i.e. $\triangle A=8\pi$. In any case, the result in \cite{key-5}
can be improved too, by taking into account the deviation from the
strictly thermal feature in Eq. (\ref{eq: Parikh Correction}), i.e.
by using Eq. (\ref{eq: quasinormal modes corrected}) than Eq. (\ref{eq: quasinormal modes}).
Assuming a transition $n\rightarrow n-1$ Eq. (\ref{eq: radice fisica})
gives an absorbed energy \begin{equation}
\triangle M=(\omega_{0})_{n}-(\omega_{0})_{n-1}=f(M,n)\label{eq: variazione}\end{equation}

where we have defined

\begin{equation}
\begin{array}{c}
f(M,n)\equiv\\
\\\equiv\sqrt{M^{2}-\frac{1}{4\pi}\sqrt{(\ln3)^{2}+4\pi^{2}(n-\frac{1}{2})^{2}}}-\sqrt{M^{2}-\frac{1}{4\pi}\sqrt{(\ln3)^{2}+4\pi^{2}(n+\frac{1}{2})^{2}}}.\end{array}\label{eq: f(M,n)}\end{equation}

Therefore

\begin{equation}
\triangle A=32\pi M\triangle M=32\pi M\times f(M,n)\label{eq: area quantum}\end{equation}

For very large $n$ one gets 

\begin{equation}
\begin{array}{c}
f(M,n)\approx\\
\\\approx\sqrt{M^{2}-\frac{1}{2}(n-\frac{1}{2})}-\sqrt{M^{2}-\frac{1}{2}(n+\frac{1}{2})}\approx\frac{1}{4M},\end{array}\label{eq: circa}\end{equation}

and Eq. (\ref{eq: area quantum}) becomes $\triangle A\approx8\pi$
which is the original result of Bekenstein for the area quantization
\cite{key-10}. Then, only in the limit $n\rightarrow\infty$ the
levels are equally spaced. Indeed, for finite $n$ there are deviations,
see Eq. (\ref{eq: non equalmente spaziati}).

This analysis will have important consequences on entropy and microstates.

Assuming that, for large $n$, the horizon area is quantized \cite{key-5}
with a quantum $\triangle A=\alpha,$ where $\alpha=32\pi M\cdot f(M,n)$
for us, $\alpha=8\pi$ for Bekenstein \cite{key-10} and Maggiore
\cite{key-5}, $\alpha=4\ln3$ for Hod \cite{key-11}, the total horizon
area must be $A=N\triangle A=N\alpha$ (notice that the number of
quanta of area, the integer $N,$ is \emph{not} the overtone number
$n$). Our approach gives:

\begin{equation}
N=\frac{A}{\triangle A}=\frac{16\pi M^{2}}{\alpha}=\frac{16\pi M^{2}}{32\pi M\cdot f(M,n)}=\frac{M}{2f(M,n)}.\label{eq: N}\end{equation}

Hawking radiation and black hole's entropy are the two most important
predictions of a yet unknown unitary quantum theory of gravity. The
famous formula of Bekenstein-Hawking entropy \cite{key-1,key-13,key-14}
now reads

\begin{equation}
S_{BH}=\frac{A}{4}=8\pi NM\triangle M=8\pi NM\cdot f(M,n),\label{eq: Bekenstein-Hawking}\end{equation}

becoming a function of the overtone number $n$.

In the limit $n\rightarrow\infty$ $f(M,n)\rightarrow\frac{1}{4M}$
and we re-obtain the standard result \cite{key-5,key-15,key-16,key-17}

\begin{equation}
S_{BH}\rightarrow2\pi N.\label{eq: entropia circa}\end{equation}

In any case, it is a general belief that here is no reason to expect
that Bekenstein-Hawking entropy to be the whole answer for a correct
theory of quantum gravity \cite{key-18}. In order to have a better
understanding of black hole's entropy, it is imperative to go beyond
Bekenstein-Hawking entropy and identify the sub-leading corrections
\cite{key-18}. In \cite{key-19} Zhang used the quantum tunnelling
approach in \cite{key-2,key-3} to obtain the sub-leading corrections
to the second order approximation. In that approach, the black hole's
entropy contains three parts: the usual Bekenstein-Hawking entropy,
the logarithmic term and the inverse area term \cite{key-19}

\begin{equation}
S_{total}=S_{BH}-\ln S_{BH}+\frac{3}{2A}.\label{eq: entropia totale}\end{equation}

The logarithmic and inverse area terms are the consequence of requesting
to satisfying the unitary quantum gravity theory \cite{key-19}. Apart
from a coefficient, this correction to the black hole's entropy is
consistent with the one of loop quantum gravity \cite{key-19}. In
fact, in loop quantum gravity the coefficient of the logarithmic term
has been rigorously fixed at $\frac{1}{2}$ \cite{key-19,key-20}.
By using the correction (\ref{eq: Bekenstein-Hawking}) to Bekenstein-Hawking
entropy Eq. (\ref{eq: entropia totale}) can be rewritten as

\begin{equation}
S_{total}=8\pi NM\cdot f(M,n)-\ln8\pi NM\cdot f(M,n)+\frac{3}{64\pi NM\cdot f(M,n)}\label{eq: entropia totale 2}\end{equation}

that in the limit $n\rightarrow\infty$ becomes

\begin{equation}
S_{total}\rightarrow2\pi N-\ln2\pi N+\frac{3}{16\pi N}.\label{eq: entropia totale approssimata}\end{equation}

Our results imply that at level $N$ the black hole has a number of
microstates

\begin{equation}
g(N)\propto\exp\left[8\pi NM\cdot f(M,n)-\ln8\pi NM\cdot f(M,n)+\frac{3}{64\pi NM\cdot f(M,n)}\right],\label{eq: microstati}\end{equation}

that in the limit $n\rightarrow\infty$ reads

\begin{equation}
g(N)\propto\exp(2\pi N-\ln2\pi N+\frac{3}{16\pi N}).\label{eq: microstati circa}\end{equation}

In summary, in this work the spectrum of black hole's quasinormal
modes has been re-analysed by taking into account, through the introduction
of an effective temperature,  the correction in \cite{key-2,key-3}
which shows that the radiation spectrum cannot be strictly thermal.
This important issue modifies in an fundamental way the physical interpretation
of the black hole's spectrum, enabling a re-examination of various
results in the literature. In particular, the formula of the horizon's
area quantization and the number of quanta of area result modified
becoming functions of the quantum overtone number $n$. Hence, the
famous formula of Bekenstein-Hawking entropy its sub-leading corrections
and the number of microstates are also modified. Black hole's entropy
becomes a function of the quantum overtone number too.

The presented results are fundamental for realizing unitary quantum
gravity theory. In fact, Hawking radiation and black hole's entropy
are the two fundamental predictions of such a definitive theory and
black holes are considered theoretical laboratories for testing models
of it. Thus, black hole's quasinormal modes are the best candidates
for an interpretation in terms of quantum levels. We emphasize that
this is the first time that black hole's entropy has been directly
connected with a quantum number.

Notice that previous results in the literature are re-obtained in
the limit $n\rightarrow\infty$. This point confirms the correctness
of the analysis in this work which improves previous approximations.


\begin{thebibliography}{20}
\bibitem{key-1}S. W. Hawking, Commun. Math. Phys. 43, 199 (1975).

\bibitem{key-2}M. K. Parikh, Gen. Rel. Grav. 36, 2419 (2004, First
Award at Gravity Research Foundation).

\bibitem{key-3}M. K. Parikh and F. Wilczek, Phys. Rev. Lett. 85,
5042 (2000).

\bibitem{key-4}T. Padmanabhan, Class. Quant. Grav. 21, L1 (2004).

\bibitem{key-5}M. Maggiore, Phys. Rev. Lett. 100, 141301 (2008).

\bibitem{key-6}H. P. Nollert, Phys. Rev. D 47, 5253 (1993).

\bibitem{key-7}N. Andersson, Class. Quant. Grav. 10, L61 (1993).

\bibitem{key-8}L. Motl, Adv. Theor. Math. Phys. 6, 1135 (2003).

\bibitem{key-9}L. Motl and A. Neitzke, Adv. Theor. Math. Phys. 7,
307 (2003).

\bibitem[10]{key-10}J. D. Bekenstein, Lett. Nuovo Cim. 11, 467 (1974).

\bibitem[11]{key-11}S. Hod, Phys. Rev. Lett. 81 4293 (1998).

\bibitem[12]{key-12}O. Dreyer, Phys. Rev. Lett. 90, 081301 (2003).

\bibitem[13]{key-13}J. D. Bekenstein, Nuovo Cim. Lett. 4, 737 (1972). 

\bibitem[14]{key-14}J. D. Bekenstein, Phys. Rev. D7, 2333 (1973). 

\bibitem[15]{key-15}A. Barvinsky and G. Kunstatter, arXiv:gr-qc/9607030.

\bibitem[16]{key-16}A. Barvinsky S. Das and G. Kunstatter, Class.
Quant. Grav. 18, 4845 (2001).

\bibitem[17]{key-17}D. Kothawala, T. Padmanabhan, S. Sarkar, Phys.
Rev. D 78, 104018 (2008).

\bibitem[18]{key-18}S. Shankaranarayanan, Mod. Phys. Lett. A 23,
1975-1980 (2008).

\bibitem[19]{key-19}J. Zhang, Phys. Lett. B 668, 353-356 (2008).

\bibitem[20]{key-20}A. Ghosh, P. Mitra, Phys. Rev. D 71, 027502 (2005). 
\end{thebibliography}
\end{document}